\documentclass[aps,prb,superscriptaddress,preprint]{revtex4-1}

\usepackage{amssymb}
\usepackage{amsmath}
\usepackage{graphicx}
\usepackage{color}

\begin{document}

\title{Conductance oscillations in Chern insulator junctions: valley-isospin dependence and Aharonov-Bohm effects }

\author{Nojoon Myoung}
\affiliation{Department of Physics Education, Chosun University, Gwangju 61452, Republic of Korea}
\affiliation{Center for Theoretical Physics of Complex Systems, Institute for Basic Science, Daejeon 34051, Republic of Korea}
\author{Hee Chul Park}
\email{hcpark@ibs.re.kr}
\affiliation{Center for Theoretical Physics of Complex Systems, Institute for Basic Science, Daejeon 34051, Republic of Korea}

\begin{abstract}
The transport properties of Chern insulator junctions generated by bipolar junctions in quantum Hall graphene are theoretically studied in the coherent regime. Coherent transport across the junction exhibits two mesoscopic features: valley-isospin dependence of the quantum Hall conductance, and the Aharonov-Bohm (AB) effects with the interface channels. We demonstrate that the valley-isospin dependence can be measured in a graphene sample with perfect edge terminations, resulting in conductance oscillation for the smallest Chern number case. On the other hand, while conductance plateaus are found to be unclear for larger Chern numbers, the conductance exhibits an oscillatory behavior of which period is relatively longer than the valley-isospin dependent oscillation. This conductance oscillation is ascribed to the AB effect, which is implicitly created by the split metallic channels near the junction interface. We point out that a possible origin of the unclear plateaus previously speculated to be incompleteness in realistic devices is the low-visibility conductance oscillation due to unequal beam splitting.
\end{abstract} 

\date{\today}

\maketitle

\section{Introduction}

Graphene is a promising material for studying quantum Hall effects with gate-tunable filling factors on account of its capability for controlling charge density via field effects\cite{Novoselov2005,CastroNeto2009}. Studies involving conductance measurements through graphene under a homogeneous magnetic field report non-integer conductance plateaus for gate-tunable bipolar junctions\cite{Williams2007,Abanin2007,Shytov2007,Kolasinska2016,Kolasinski2017,Sekera2017,Handschin2017}. The topological nature of the quantum Hall system has been clearly understood via the presence of gapless edge states that are topologically protected\cite{Hatsugai1993,Kane2005}. Bulk-boundary correspondence offers an intuitive way of understanding the properties of these edge states: the number of conducting channels is characterized by the topological invariant of the quantum Hall insulator\cite{Halperin1982,Essin2011,Matsuura2013}. It has been well known that the topological invariant (or so-called Chern number) of a quantum Hall insulator is given by the filling factor in the integer quantum Hall effect\cite{Thouless1982,Haldane1988,Hasan2010,Ando2013}.

The observation of non-integer conductance plateaus in bipolar graphene quantum Hall systems has been interpreted by the equilibration of interface states at the p-n junction, with theoretical efforts supporting experimental findings by considering edge and interface disorders\cite{Li2008,Low2009,Frassdorf2016,LaGasse2016}. Junction conductance via interface equilibration has also been reported for p-n-p junctions in quantum Hall graphene systems\cite{Ki2009,Ki2010,Debey2016}, with the consideration that there can be reflections at the bipolar junction. These studies were carried out in macroscopic systems where mesoscopic fluctuations were ignored\cite{Abanin2007}. However, T. Low has shown that observed junction conductance in ballistic systems is distinct from disordered ones, via crossover between the coherent and Ohmic regimes\cite{Low2009}. Mesoscopic conductance fluctuation should therefore be expected to appear in the coherent regime, e.g., a valley-isospin dependence of the quantum Hall effects in graphene p-n junctions\cite{Tworzydlo2007}.

In this paper, we show that mesoscopic consequences in the conductance across a Chern insulator junction can be observed even in the presence of edge disorders, when both regions of the Chern insulator junction are on the second Hall plateaus. As the length of the junction interface varies, we reveal that the conductance across the junction exhibits atomic-scale period fluctuation and long-period oscillation according to the Chern number configuration. While this fluctuation, associated with valley-isospin dependence, can be eliminated by the presence of edge roughness, the long-range conductance oscillation survives despite a randomly distributed edge roughness. We demonstrate that the conductance oscillation originates from an Aharonov-Bohm (AB) interferometry implicitly contained in the Chern insulator junction; since the metallic channels around the interface are spatially separated, they effectively create an area enclosing magnetic flux. The AB conductance oscillation also exhibits a beating pattern with a very long period, reflecting the multi-path interferometry of the implicit AB ring. Finally, we discover a gate-tunable visibility of the AB oscillation and further show that a suppression of the AB conductance oscillation can be achieved through gate control.

Our paper is organized as follows. In Sec. \ref{sec:model}, we give an account of our theoretical formalism.  We discuss the conductance spectra through the Chern insulator junction in Sec. \ref{sec:cond} by considering different Chern number configurations. In Sec. \ref{sec:isospin}, we investigate the effects of edge roughness on the valley-isospin dependence. Sec. \ref{sec:ABosc} presents our interesting finding that the conductance oscillation occurs due to the AB effect intrinsic to the single Chern insulator junction. We discuss the properties of the intrinsic AB interferometry in Sec. \ref{sec:suppAB}, which are expected to be of interest to practical device fabrication and measurements, and conclude in Sec. \ref{sec:conclusion}.

\section{Theoretical approaches to the Chern insulator junction} \label{sec:model}

\begin{figure}[hbpt!]
\includegraphics[width=8.5cm]{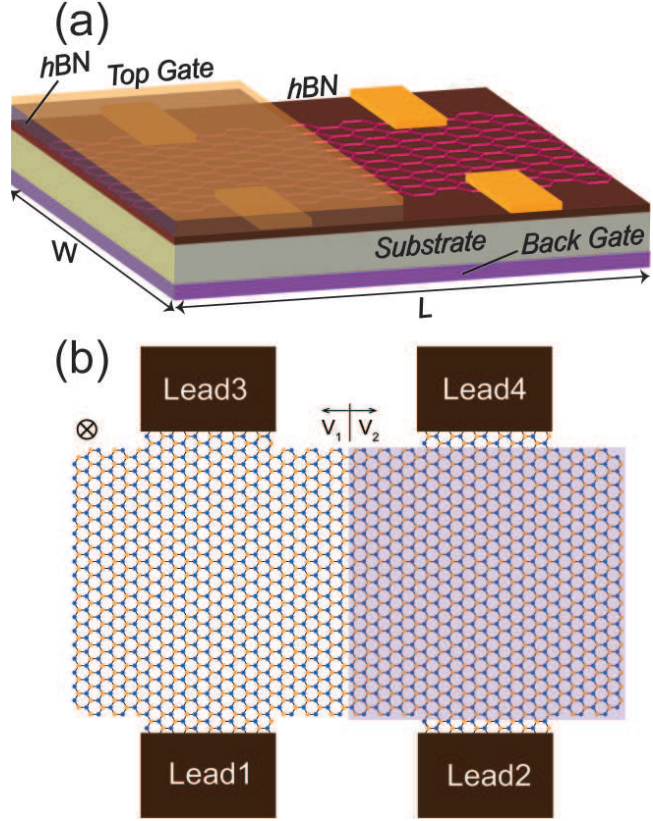}
\caption{(a) Schematic diagram of the gated structure considered in this study. Graphene is encapsulated by h-BN layers. (b) Depiction of the four-terminal graphene Hall bar where a Chern insulator junction is created by exploiting the above gated structure.} \label{fg:repfigs}
\end{figure}

In terms of topology, quantum Hall states have been revealed to have topological characteristics and can regarded as Chern insulator\cite{Thouless1982,Hatsugai1993}. Since Chern numbers of quantum Hall states in graphene are tunable by electric-field effect, a heterojunction of different Chern insulators is expected to be realized by using the bipolar junction of graphene in the quantum Hall regime\cite{Liu2015}. Such a device structure is feasible to fabricate with gated structures under a homogeneous magnetic field\cite{Williams2007,Ozylmaz2007}(see Fig. \ref{fg:repfigs}). Let us note that there should be a thin dielectric layer (e.g. few-layer h-BN) between graphene and the top gate electrode, although we omit it here for simplicity. With an analytical approach, the effective Dirac Hamiltonian for graphene under a homogeneous magnetic field reads
\begin{align}
H=\hbar v_{F}\vec{\sigma}\cdot\vec{\pi}+V\left(x\right),
\end{align}
where $v_{F}\simeq10^{6}$ ms$^{-1}$ is the Fermi velocity of graphene, $\vec{\sigma}=\left(\sigma_{1},\sigma_{2}\right)$ are the Pauli matrices, $\vec{\pi}=\vec{p}+e\vec{A}$, and the electrostatic potential is given as either a sign function or hyperbolic tangent function:
\begin{subequations}
\begin{align}
V\left(x\right)&=V_{0}~\mbox{sign}\left(x\right),\label{eq:bipolarcontinuum_a}\\
V\left(x\right)&=V_{0}\tanh{\left(\frac{x}{\xi}\right)}, \label{eq:bipolarcontinuum}
\end{align}
\end{subequations}
where $V_{0}=V_{2}=-V_{1}$; $V_{1}$ and $V_{2}$ are potential energies in the left and right sides of the p-n junction\cite{Wang2012}. Equation (\ref{eq:bipolarcontinuum_a}) represents an abrupt step, and Eq. (\ref{eq:bipolarcontinuum}) indicates a smoothly varying step for $\xi\ne0$; the latter case will be considered here for discussion on the effects of a smooth junction. Note that for the limit $\xi\rightarrow0$, Eq. (\ref{eq:bipolarcontinuum}) approximates to Eq. (\ref{eq:bipolarcontinuum_a}). The Dirac equation with the above effective Hamiltonian, $H\Psi=E\Psi$ with $\Psi=\left(\psi_{A},\psi_{B}\right)^{T}$, consists of two sublattice-coupled equations and becomes analytically solvable by decoupling them, when we consider the aburpt potential step, resulting in the following Schr\"{o}dinger-like second-order differential equation\cite{DeMartino2007,Ghosh2008,Masir2008,Kuru2009,Myoung2011,Downing2016} (details are provided in Appendix \ref{sec:decouple}):
\begin{align}
\left[\frac{d^{2}}{dx^{2}}+\frac{\varsigma}{2}-\left(k_{y}-\frac{x}{2}\right)^{2}+\left(E-V\right)^{2}\right]\psi_{A,B}=0, \label{eq:effHam}
\end{align}
where $\varsigma=\pm1$ for different sublattices (A and B), $V$ is a electric potential in each region of the p-n junction, and $k_{y}$ is the $y$-component momentum which acts as a good quantum number since $\left[H,p_{y}\right]=0$. The above equation is dimensionless upon $E_{0}=\sqrt{2\hbar v_{F}^{2}eB}$ and $l_{B}=\sqrt{\hbar/\left(2eB\right)}$, and we choose the Landau gauge, i.e., $\vec{A}=\left(0,-Bx,0\right)$, which leads to $\vec{B}=\left(0,0,-B\right)$. Note that we have $E_{0}\simeq200$ meV and $l_{B}\simeq3.3$ nm for $B=30$ T. Equation (\ref{eq:effHam}) can be regarded as if Dirac fermions experience the effective potential given by
\begin{align}
V_{eff}=-\frac{\varsigma}{2}+\left(k_{y}-\frac{x}{2}\right)^{2}, \label{eq:effpot}
\end{align}
which is valid for the abrupt potential step case. The solutions for $\psi_{A}$ and $\psi_{B}$ are obtained as parabolic cylinder functions defined by Whittaker and Watson\cite{Whittaker1990}, in the low-energy limit\cite{Oroszlany2008,Park2008,Ghosh2008,Cohnitz2016}:
\begin{align}
\Psi\left(x\right)=\left(\begin{array}{c}\psi_{A}\left(x\right)\\ \psi_{B}\left(x\right)\end{array}\right)=A\left(\begin{array}{c}D_{\nu}\left(s\zeta\right)\\ -\mathrm{i}s\sqrt{\frac{\nu}{2}}D_{\nu-1}\left(s\zeta\right)\end{array}\right), \label{eq:analsols}
\end{align}
where $\nu\equiv\left(E-sV_{0}\right)^{2}$ for each region of the Chern insulator junction, $\zeta\equiv 2k_{y}-x$, and $s=sign\left(x\right)$. Note that the normalization factor is found to be $A=\left[16\pi\left(\nu-1\right)!^{2}\right]^{1/4}$.

Alternatively, the tight-binding approach also accounts for the present system, leading to the following Hamiltonian:
\begin{align}
H=\sum_{i}\epsilon_{i}c_{i}^{\dagger}c_{i}+\sum_{\left\langle i,j\right\rangle}t_{ij}\left(c_{i}^{\dagger}c_{j}+h.c.\right),
\end{align}
where $\epsilon_{i}$ is the on-site energy corresponding to the potential step (Eq. (\ref{eq:bipolarcontinuum})), and $c_{i}^{\dagger}$ and $c_{i}$ are the creation and annihilation operators on the $i$-th site. In the presence of a magnetic field, the hopping term is defined by
\begin{align}
t_{ij}=t e^{\mathrm{i}\frac{2\pi}{\Phi_{0}}\int_{r_{i}}^{r_{j}}\vec{A}\cdot d\vec{r}},
\end{align}
where $t$ = 3.0 eV is the hopping energy, and $\Phi_{0}=e/h$ is the flux quantum. Here, we choose the same gauge considered in the analytical approach. Note that the system is centered at $\vec{r}=\left(0,0\right)$. Since we introduce a p-n junction to graphene through the potential Eq. (\ref{eq:bipolarcontinuum}), the system is divided into n- and p-doped regions where the zeroth Landau levels (LLs) are located at $E=-V_{0}$ and $V_{0}$, respectively. Ballistic conductance of the four-terminal graphene Hall bar is calculated in the linear response regime, exploiting the Landauer-B\"{u}ttiker approach, as
\begin{align}
G_{\alpha\beta}\left(E\right)=\frac{e^{2}}{h}\sum_{a\in\alpha,b\in\beta}\left|S_{ab}\left(E\right)\right|^{2},
\end{align}
where $S_{ab}$ is the scattering matrix from channels $b$ to $a$, which belong to lead $\beta$ and $\alpha$, respectively. With the tight-binding Hamiltonian, we can also take local density of states (LDOS) and the probability density in the scattering region for the incoming wave through a given lead at a given energy by using \textsc{KWANT} packages\cite{Groth2014}. Lastly, note that every attached lead is semi-infinitely long in the numerical calculations with translational symmetry based on the translational vectors of graphene, even though they are not displayed in the map figures for the numerical results of LDOS and wavefunctions in this paper.

\begin{figure}[hbpt!]
\includegraphics[width=8.5cm]{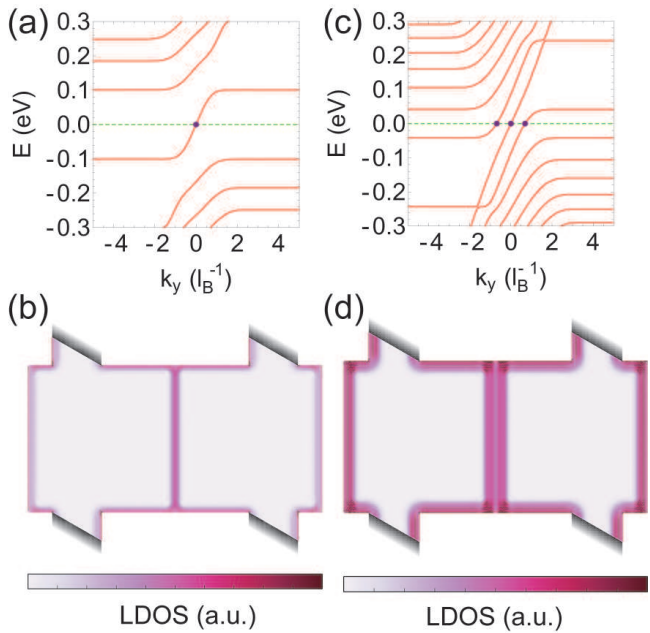}
\caption{(a) and (c) Eigenenergies of the system with Chern insulator junctions $V_{0}=E_{0}/2$ and $\left(\sqrt{2}+1\right)E_{0}/2$ at $B=30$ T, respectively. The dashed lines represent the given Fermi energy $E_{F}=0$ and the dots mark the intersections of the Fermi energy and the eigenenergies, indicating the existence of interface states. (b) and (d) Local densities of states corresponding to (a) and (c), respectively. The light and dark colors represent low and high densities.} \label{fg:repfigs2}
\end{figure}

In this study, numerical results from the tight-binding calculations are qualitatively interpreted through the analytic approach. For the given potential $V_{0}=E_{0}/2$, the existence of a metallic channel at the p-n junction interface is observed by the LDOS at $E_{F}=0$ with the numerical calculations, as shown in Fig. \ref{fg:repfigs2}(c). In order to understand the origin of the metallic channel, one can see eigenenergy bands at the interface through the use of wavefunction continuity conditions at $x=0$ for the same potential (details are provided in Appendix \ref{sec:eveq}.) Figure \ref{fg:repfigs2}(a) clearly shows that a metallic channel must exist at the interface, connecting the same-index LLs. Similarly, Fig. \ref{fg:repfigs2}(b) also supports the existence of the metallic channel shown in Fig. \ref{fg:repfigs2}(d) for $V_{0}=\left(\sqrt{2}+1\right)E_{0}/2$, where we see a larger number of channels than in the previous case. Notice that the eigenenergy bands (Figs. \ref{fg:repfigs2}(a) and (b)) reflect the presence of electronic states at the interface, and so differ from the eigenenergies of the whole finite graphene system.

It is worth mentioning that the occurrence of the metallic channels along the Chern insulator junction is due to the bulk-boundary correspondence; therefore, the interface states are topologically protected. The number of interface states is in keeping with the Chern number configuration of the quantum Hall graphene system. In fact, the Chern number of the quantum Hall insulator is derived from the TKNN formalism\cite{Thouless1982,Hatsugai1993}, and the Chern number of each region turns out to be the filling factor, i.e., $\mathcal{C}\times2e^{2}/h\equiv\nu\times2e^{2}/h$ where $\mathcal{C}$ and $\nu$ represent the Chern number and filling factor. A similar phenomenon was also reported in a heterotype Chern insulator junction created by using inhomogeneous magnetic fields\cite{Liu2015}. Note that, here, we define the odd-number filling factor of the graphene quantum Hall regime in units of $2e^{2}/h$, distinct from the even-number filling factor in units of $e^{2}/h$. Since the interface states are doubly degenerate on account of valley symmetry, the single channel at the interface in both Figs. \ref{fg:repfigs2}(a) and (b) actually contains two interface states, and the number of the interface states equals the Chern number difference for $\left(\mathcal{C}_{1},\mathcal{C}_{2}\right)=\left(1,-1\right)$. For Figs. \ref{fg:repfigs2}(c) and (d), the Chern number configuration is defined as $\left(\mathcal{C}_{1},\mathcal{C}_{2}\right)=\left(3,-3\right)$, where both regions are on the first quantum Hall plateau, i.e., an insulating phase between the zeroth and first LLs.

\section{Four-terminal conductance in Chern insulator junctions} \label{sec:cond}

As reported in previous studies\cite{Williams2007,Abanin2007}, the interface states in bipolar quantum Hall graphene exist only when the Chern numbers on each side of the Chern insulator junctions are opposite, on the basis of the equilibration concept. Since the system is finite, the junction interface meets two ends at its boundary: the parallel-propagating states are mixed at the bottom of the interface, and the mixed states are split into opposite directions along the system boundaries at the top. The splitting may result in reflected modes at the interface, which have been indirectly measured in multi-terminal devices\cite{Ki2009,Ki2010}. Likewise, using four-terminal geometry, we are able to individually measure the splitting of the conductance through the Chern insulator junction with two transverse conductances, $G_{31}$ and $G_{41}$, which are taken between leads 1 and 3 or 4, as displayed in Fig. \ref{fg:repfigs}(b). In addition, we also take into account the longitudinal conductance $G_{21}$ to detect the edge state contribution to the conductance for the Chern insulator junctions with the same sign of Chern numbers.

\begin{figure*}[hbpt!]
\includegraphics[width=16cm]{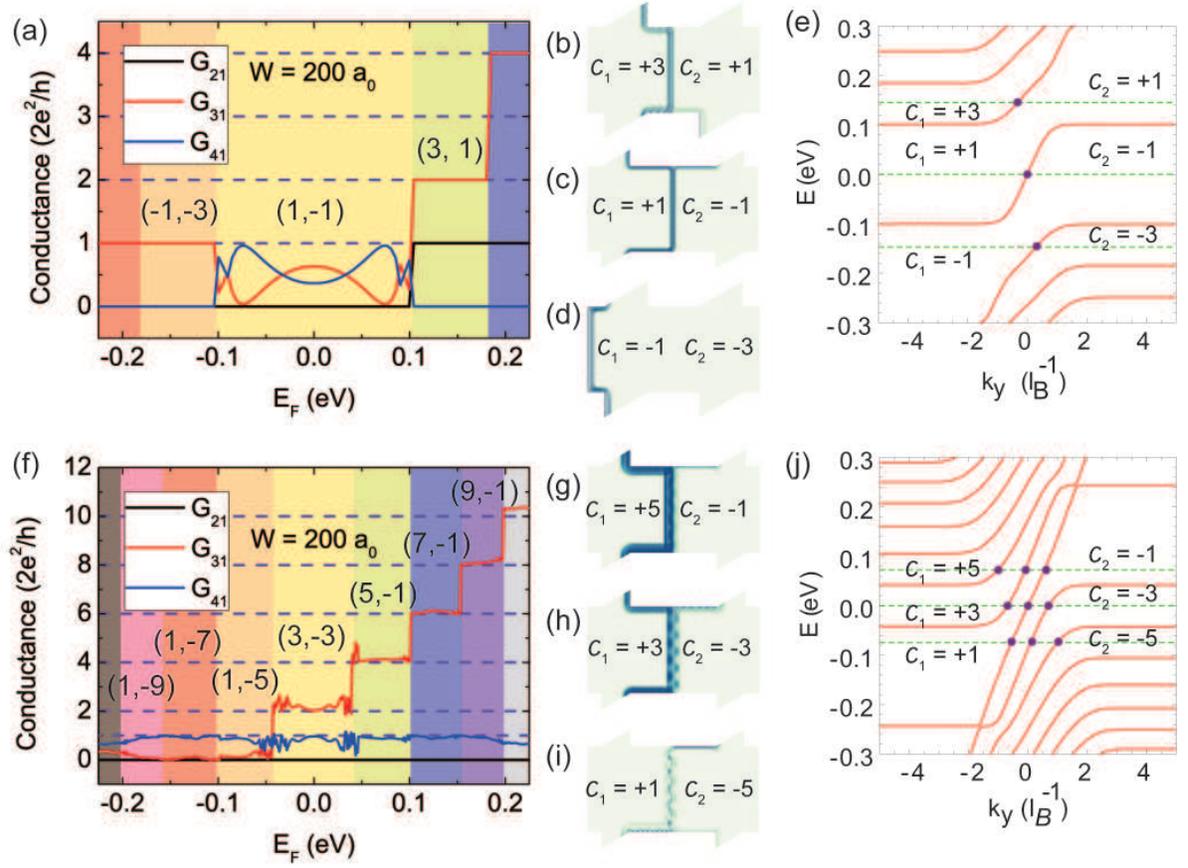}
\caption{ (a) and (f) Four-terminal conductance through the Chern insulator junctions produced in the graphene quantum Hall regime for given potentials $V_{0}=E_{0}/2$ and $V_{0}=\left(\sqrt{2}+1\right)E_{0}/2$, respectively. Colored regions represent various Chern insulator junction cases with different Chern number configurations, $\left(\mathcal{C}_{1},\mathcal{C}_{2}\right)$, depending on the Fermi energy. The dashed lines display quantized conductance values. (b), (c), and (d) Probability densities of the propagating modes, coming in from lead 1, at Fermi energies $E_{F}=E_{0}/\sqrt{2}$, 0, and $-E_{0}/\sqrt{2}$, respectively. (e) Eigenenergies of the Chern insulator junction in (a). From the top down, each dashed line indicates the given Fermi energy corresponding to (b), (c), and (d), respectively, and the dots on the intersections of the eigenenergy bands with the Fermi energies imply the existence of the interface states. (g), (h), and (i) Probability densities at Fermi energies $E_{F}=\left(\sqrt{3}-1\right)E_{0}/2$, 0, and $-\left(\sqrt{3}-1\right)E_{0}/2$, respectively. (j) Eigenenergies of the Chern insulator in (f). From the top down, each dashed line indicates the given Fermi energy corresponding to (g), (h), and (i), respectively. } \label{fg:cleancond}
\end{figure*}

Figure \ref{fg:cleancond}(a) shows the numerically calculated conductances for the given Chern insulator junction as functions of Fermi energy. This case considers the simplest junction where a metallic channel at the interface is formed between $\mathcal{C}_{1}=+1$ and $\mathcal{C}_{2}=-1$ regions as shown in Fig. \ref{fg:cleancond}(e). This junction is created by applying the potential $V_{0}=E_{0}/2$. When $\left|E_{F}\right|<E_{0}/2$, each region of the Chern insulator junction delivers electron- and hole-like edge modes with opposite directions of circulation, and a metallic channel at the interface results from the mixing of the electron- and hole-like modes. It is seen that $G_{21}$ vanishes for $\left|E_{F}\right|<E_{0}/2$ because the incoming mode wholly propagates along the interface, as shown in Fig. \ref{fg:cleancond}(c). The existence of the interface channel leads to non-zero $G_{31}$ and $G_{41}$, which are attributed to beam splitting via the interface channel (see Fig. \ref{fg:cleancond}(c)). One can clearly see that through flux conservation, the sum of $G_{31}$ and $G_{41}$ always becomes $2e^{2}/h$, i.e., $\mathcal{C}_{1}G_{0}$, which is equal to the quantum Hall conductance plateau from the zeroth-LL state\cite{Novoselov2005}.

On the other hand, for $E_{F}<-E_{0}/2$, both $\mathcal{C}_{1}=-1$ and $\mathcal{C}_{1}=-3$ regions deliver hole-like edge modes with the same circulating direction, causing the cancellation of the two counter-propagating modes at the interface. Because of the absence of the interface mode, it is obvious to see $G_{41}=0$, and the reversed circulating direction in the $\mathcal{C}_{1}=-1$ region allows for $G_{31}=2e^{2}/h$, attributed to a flux flow along the boundary (Fig. \ref{fg:cleancond}(d)).

Lastly, for $E_{F}>E_{0}/2$, both $\mathcal{C}_{1}=+3$ and $\mathcal{C}_{1}=+1$ regions give rise to electron-like edge modes with the same circulating direction. It is straightforward to expect that the edge modes are cancelled out at the interface due to the same direction of circulation in both regions, leaving one metallic channel in the $\mathcal{C}_{1}=+3$ region, as displayed in Fig. \ref{fg:cleancond}(b). Similar to the $\left(-3,-1\right)$ case, the cancelled interface channel prevents the incoming wave from propagating through the interface ($G_{41}=0$); rather, it wholly propagates along the boundary and results in a $G_{21}$ plateau at $2e^{2}/h$. The absence of the metallic channel formed at the exact interface allows for the stepwise increase in $G_{31}$ as a function of $E_{F}$, attributed to the spatially shifted metallic channel from the junction interface. It is noteworthy here that the position of the interface channels is changeable, depending on the Fermi energy. As displayed in Fig. \ref{fg:cleancond}(d), the metallic channels at $E_{F}=0$ are found to be for $k_{y}=0$, but at $E_{F}\neq0$, the interface states are found to be for nonzero $k_{y}$. The position of the metallic channels is easily understood by the effective potential $V_{eff}$ (Eq. (\ref{eq:effpot})), which is regarded as a harmonic potential centered at $2k_{y}$. Therefore, we can discern the position of the metallic channels at $x=2k_{y}$ where $k_{y}$ is obtained by finding intersections between the given Fermi energy and the eigenenergy bands.

Next, we consider the more complicated case, where the Chern number configuration is given as $\left(3,-3\right)$ by applying the potential $V_{0}=\left(\sqrt{2}+1\right)E_{0}/2$. Figure \ref{fg:cleancond} (f) shows the numerically calculated results for the four-terminal conductances as functions of Fermi energy. In this case, because the potential difference between two regions is large enough, the Chern numbers of each region have the opposite sign; i.e., there exists one metallic channel at the exact interface in the given range of Fermi energy, as depicted in Fig. \ref{fg:cleancond}(f)). The transport phenomena in this case are similar to the simplest cases: i) one metallic channel is always formed at the exact interface for the opposite-sign Chern number configurations, causing the split conductances $G_{31}$ and $G_{41}$ but prohibiting flux flow between leads 1 and 2 $\left(G_{21}=0\right)$; and ii) other channels are generated with distances from the interface that exhibit a stepwise increase in $G_{31}$ like typical quantum Hall effects. Here, it is worthwhile to mention that $G_{31}+G_{41}=C_{1}G_{0}$, due to the flux conservation.

Figures \ref{fg:cleancond}(g), (h), and (i) display the transport phenomena and the formation of the metallic channels either at the exact interface or nearby the interface, supported by the eigenenergy spectra shown in Fig. \ref{fg:cleancond}(j). However, one can definitely notice that the conductance spectra do not exhibit clear plateaus, contrary to the simplest Chern insulator junction. Although such unclear conductance plateaus have been observed and understood by disorder-induced mode mixing at rough edges\cite{Williams2007,Abanin2007,Low2009,LaGasse2016}, as we consider the coherent regime for the calculations here there are no incoherent consequences related to disorders normally expected to exist in macroscopic systems. Therefore, we deduce that there must be mesoscopic consequences affecting the vagueness of the conductance plateaus and causing their obscurity, even in the coherent regime.

\subsection{Valley-isospin-dependent conductance fluctuation} \label{sec:isospin}

\begin{figure*}[hptb!]
\includegraphics[width=16cm]{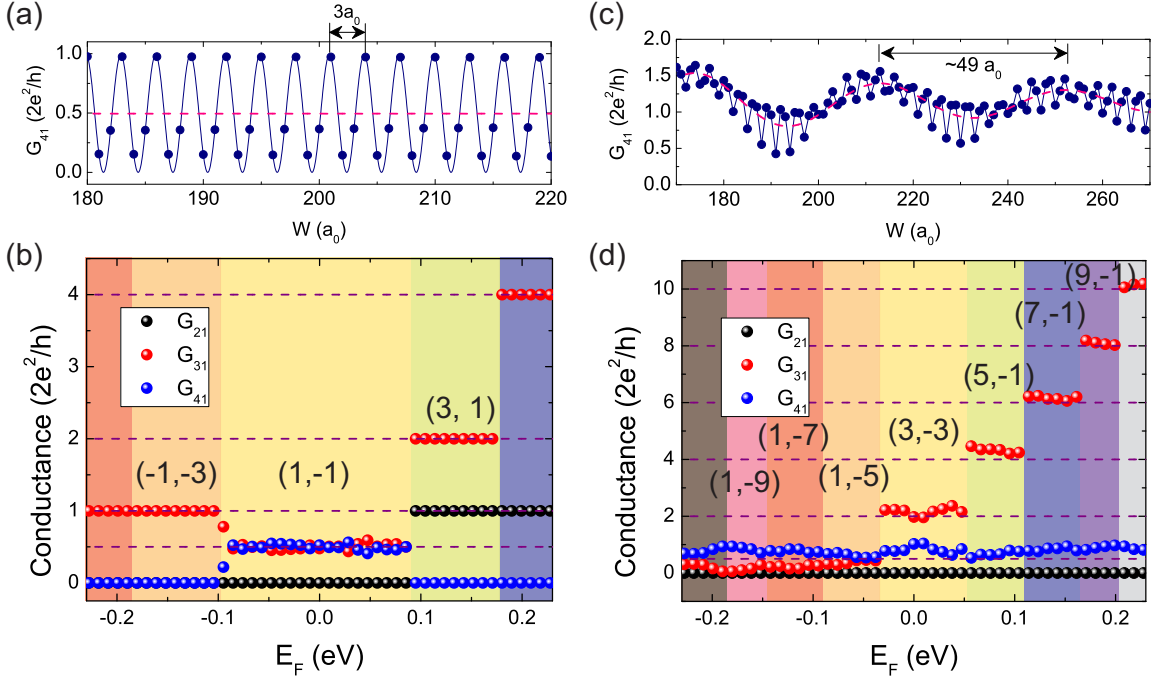}
\caption{(a) and (c) Zero-energy conductance across the junction as a function of system size for given potentials $V_{0}=E_{0}/2$ and $\left(\sqrt{2}+1\right)E_{0}/2$, respectively. Dots represent the conductance values and dashed lines indicate the average value of three adjacent data. The solid line in (a) is the fitted curve from the calculated data, and in (c) connects adjacent dots as an eye guide. (b) and (d) The averaged four-terminal conductances with the randomly distributed edge roughness as a function of Fermi energy for given potentials $V_{0}=E_{0}/2$ and $\left(\sqrt{2}+1\right)E_{0}/2$, respectively. Results are obtained by averaging over 100 individual sets with the random roughness at both edges. Colored regions represent different Chern number configurations $\left(\mathcal{C}_{1},\mathcal{C}_{2}\right)$.} \label{fg:roughcond}
\end{figure*}

The results of the four-terminal conductances for both cases in Fig. \ref{fg:cleancond} have been produced for given system size $W=200~a_{0}$. It has been theoretically revealed that the conductance values across a p-n junction in graphene largely depend on system size according to the edge terminations\cite{Tworzydlo2007}. In particular, when the edges perpendicular to the junction interface are terminated in an armchair shape, the conductance plateau varies depending on the angle between the valley isospins on each edge. As displayed in Fig. \ref{fg:repfigs}(b), the top and bottom edges of our system are terminated in an armchair shape. In Fig. \ref{fg:cleancond}(a), the conductance values are found to be $\sim0.35\times 2e^{2}/h$, perhaps corresponding to $0.25\times 2e^{2}/h$ as one of the three-fold conductance plateaus for valley-isospin dependence. Indeed, Fig. \ref{fg:roughcond}(a) shows that the conductance values at $E_{F}=0$ fluctuate from $\sim0.2\times 2e^{2}/h$ to $\sim 2e^{2}/h$ as $W$ varies.

The conductance fluctuation shown in Fig. \ref{fg:roughcond}(a) is a mesoscopic consequence of valley-isospin dependence at atomic-scale precision. However, to our knowledge, such an atomic-scale dependence of electronic properties on the size of graphene has not been confirmed by electrical measurements, because perfect edge terminations are necessary to expect the valley-isospin dependence in graphene nanoribbons. Thus, fabrication of ultra clean edge terminations in graphene is required to experimentally demonstrate the valley-isospin dependence---very recently, an experimental work indeed reported conductance oscillation due to valley isospin in quantum Hall graphene systems.\cite{Handschin2017}

Now, we theoretically demonstrate that the valley-isospin dependence becomes detectable only if graphene is terminated by perfect edges. $G_{41}$ values are compared between perfect and rough edge termination, with the latter introduced by a randomly distributed roughness on the edges of the system (details are provided in Appendix \ref{sec:random}). Since the edge modes of a Chern insulator are topologically protected, coherence still remains even in the presence of edge roughness. Figure \ref{fg:roughcond}(b) shows the resulting conductances, obtained by averaging over 100 individual sets of random roughness on the edges. The edge roughness does not affect the conductance spectra for $\left|E_{F}\right|<E_{0}/2$, but the split conductances $G_{31}$ and $G_{41}$ now exhibit clear plateaus at $e^{2}/h$, for $\left|E_{F}\right|<E_{0}/2$. The $e^{2}/h$ conductance plateau is equal to the averaged values of the three-fold conductance plateaus provided by valley-isospin dependence\cite{Tworzydlo2007} (dashed line in Fig. \ref{fg:roughcond}(a)). By getting rid of edge roughness, $G_{41}$ becomes consistent with the three-fold values for valley-isospin dependence as $W$ varies. Thus, we can conclude that the atomic-scaled valley-isospin dependence can only be experimentally measured by reducing edge roughness in a graphene sample, which is why experimental works have not as of yet successfully measured the conductance fluctuation mediated by the valley isospins.

By increasing the potential difference $V_{0}$ between two regions of the Chern insulator junction, the conductance across the junction at $E_{F}=0$ exhibits more complicated oscillation behavior, as shown in Fig. \ref{fg:roughcond}(c). The averaged conductance over three adjacent values now oscillates with a period of $\sim49a_{0}$, contrary to the constant values in Fig. \ref{fg:roughcond}(a). The conductance fluctuation with a short period $3a_{0}$ still remains, but does not seem to be regular because the atomic-scale fluctuation is added to the background oscillation. Due to the relatively longer period of the conductance oscillation, the oscillation observed in Fig. \ref{fg:roughcond}(c) is not eliminated by the given edge roughness magnitude of $~0.53a_0$. Figure \ref{fg:roughcond}(d), even in the presence of the random roughness, indeed shows that the results of the averaged conductance spectra are almost unchanged from the results of the smooth edges in Fig. \ref{fg:cleancond}(f). In the following subsection, we concentrate on the characteristics and origin of the long-period conductance oscillation beyond the valley-isospin-associated conductance fluctuation. 

\subsection{Aharonov-Bohm conductance oscillation} \label{sec:ABosc}

\begin{figure*}[hbpt!]
\includegraphics[width=14cm]{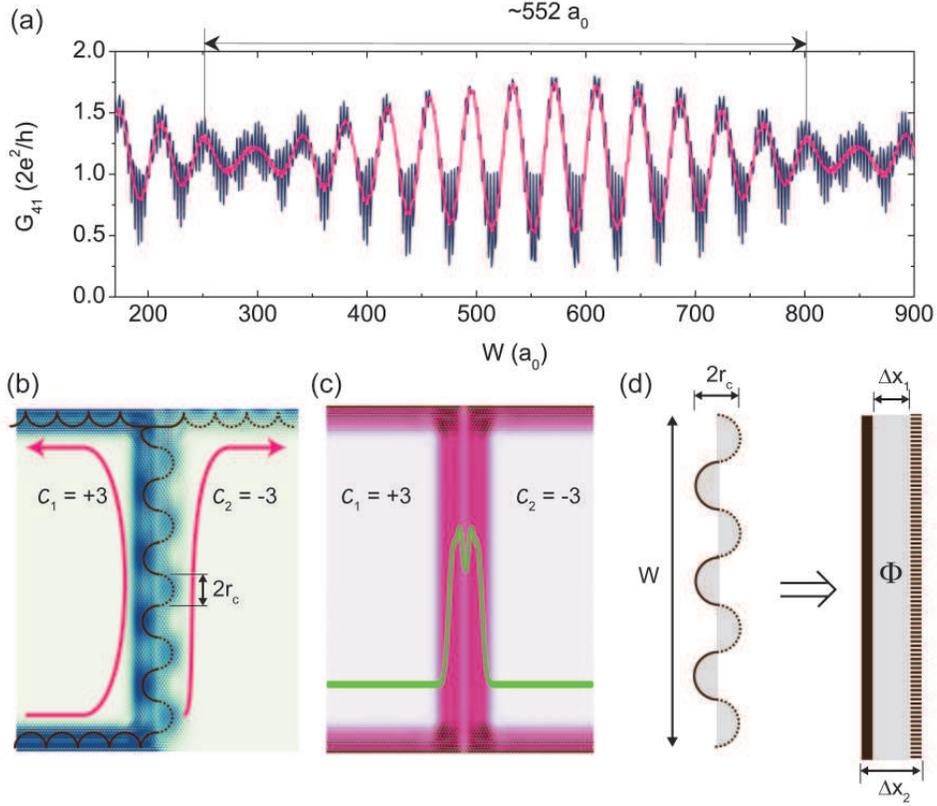}
\caption{(a) Beat of the conductance across the junction for given potential $V_{0}=\left(\sqrt{2}+1\right)E_{0}/2$ as a function of $W$. The beat period $W_{beat}$ is found to be $\sim552a_{0}$. The dark blue fluctuating line indicates the conductance values calculated from the numerical approach, and the pink oscillating line corresponds to the averaged values over three-adjacent data points showing the beat oscillation. (b) Probability density map at $E_{F}=0$ for an incoming mode from the left of the bottom edge, with illustrations of the pathway of the metallic channels and the corresponding semi-classical skipping motions. The skipping motions are characterized by the cyclotron radius $r_{c}$, depicted as solid (electron-like mode) and dashed (hole-like mode) lines. Each region has helical edge modes with oppositely circulating directions, denoted as arrows. (c) Local density of states map corresponding to (b). The absolute-squared wavefunction derived from the analytic solutions is overlaid on the map as a solid green line showing the in-level splitting of the interface states. (d) Schematics of the implicit AB interferometry. Left panel: A skipping motion along the $W$-long interface encloses an area through which magnetic fluxes penetrate. Right panel: An effective interferometry encloses the same area as that enclosed by the skipping motion, through which magnetic flux $\Phi$ penetrates in total. Both pathways are regarded as finite-width arms of the AB interferometry, characterized by the outer and inner distances $\Delta x_{1,2}$. } \label{fg:beatcond}
\end{figure*}

In order to understand the oscillating nature of the conductance for the $\left(3,-3\right)$ case, we consider a closed loop composed of spatially separated metallic channels enclosing a magnetic flux through the loop like with AB interferometry. As aforementioned, in the $\left(3,-3\right)$ case there are metallic channels formed near the interface with spatial separations; here, let us define the distance between those channels as $\Delta x$. The  conductance oscillation for AB interferometry is described by\cite{Datta1995}
\begin{align}
G_{41}\propto 1-\cos{\left(2\pi\frac{\Phi}{\Phi_{0}}+\varphi_{0}\right)},
\end{align}
where $\Phi=B\Delta xW$ is a magnetic flux enclosing the AB loop, $\Phi_{0}=h/e$ is flux quantum, and $\varphi_{0}=\pi$ is Berry's phase in graphene. From the estimated period in Fig. \ref{fg:roughcond}(c), we can approximately obtain $\Delta x=2\sqrt{2}l_{B}=2r_{c}^{rms}$, where $r_{c}=2l_{B}$ is the cyclotron radius in graphene and $r_{c}^{rms}=r_{c}/\sqrt{2}$ is the root mean square value of the skipping orbit at the interface (see Fig. \ref{fg:beatcond}(d)). Interestingly, the conductance does not exhibit the long-period oscillation in the $\left(1,-1\right)$ case, because a finite area enclosing magnetic flux cannot be created by one metallic channel formed at the exact interface. In other words, such implicit AB interferometry is expected only for larger Chern number configurations.

From Fig. \ref{fg:roughcond}(c), we can see that the AB conductance oscillates between $\approx0.7\times2e^{2}/h$ and $\approx1.6\times2e^{2}/h$, but not between 0 and $3\times2e^{2}/h$. This implies that at one node of the AB interferometry, the electron beam is not equally split, resulting in worse visibility. Indeed, in this study, beam splitting at the bottom end of the junction interface leads to unequal probability densities at each metallic channel, as displayed in Fig. \ref{fg:cleancond}(h). One can also notice that not only is the period of the oscillation slightly changed, but the oscillation amplitude is gently attenuated. Attenuation should not happen here though as we consider the fully coherent regime without any inelastic scatterings. In order to verify the behavior of the AB oscillation, we further increase $W$ and find an interesting beat with a very long period $\sim552a_{0}$ (see Fig. \ref{fg:beatcond}(a)). Such a pattern can occur when there are more than two metallic channels within the arms of an AB interferometry\cite{Huefner2010}. Similarly, we assume that there are two different loops which perhaps become inner and outer loops for the finite-width metallic channels as exhibited in Fig. \ref{fg:beatcond}(c):
\begin{align}
G_{41}&\propto A_{1}\left\{1-\cos{\left(2\pi\frac{\Phi_{1}}{\Phi_{0}}+\varphi_{0}\right)}\right\}\nonumber\\&+A_{2}\left\{1-\cos{\left(2\pi\frac{\Phi_{2}}{\Phi_{0}}+\varphi_{0}\right)}\right\},
\end{align}
where $\Phi_{1,2}=B\Delta x_{1,2}W$ are magnetic fluxes enclosed by the different loops with different $\Delta x_{1,2}$ as denoted in Fig. \ref{fg:beatcond}(d). Here, $A_{1,2}$ ($A_{1}\simeq A_{2}$) are undetermined coefficients indicating the beam splitting ratio into different paths of the AB interferometry. The beat period composed of two waves is given by $\Delta W_{beat}=\left(2\pi e B\Delta x_{1}/h+2\pi e B\Delta x_{2}/h\right)^{-1}$, where we define $\Delta x_{1}=\Delta x+\delta x/2$ and $\Delta x_{2}=\Delta x-\delta x/2$. From the acquired period in Fig. \ref{fg:beatcond}(a), we can find $\delta x\approx0.6~a_{0}$. This very small value suggests that the implicit AB interferometry created in the Chern insulator junction can offer an ultra-sensitive detector to discriminate electron path length at atomic-scale precision.

\subsection{Suppression of the AB conductance oscillation} \label{sec:suppAB}

\begin{figure}[hbpt!]
\includegraphics[width=8.5cm]{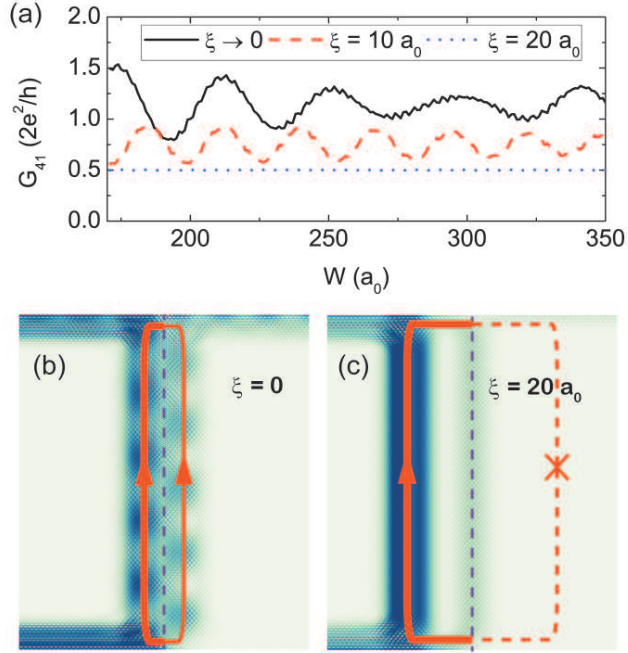}
\caption{(a) Suppression of the implicit AB oscillations as the potential profile changes. (b) and (c) Probability density maps for different potential profiles with $\xi\rightarrow0$  and 20 $a_{0}$, respectively. The closed AB interferometry loops are overlaid on the corresponding maps. The different thicknesses imply the ratio of beam splitting into each pathway, and the dashed line with an $\times$ indicates no splitting into the metallic channel in the $\mathcal{C}_{2}=3$ region. } \label{fg:dephasing}
\end{figure}

As discussed, the implicit AB interferometry in the Chern insulator junction has low values of visibility ($\approx0.5$ at maximum around the peak of the beat conductance oscillation) because the splitting of the incoming mode at the bottom of the junction interface is not half-and-half. In fact, the degree of beam splitting is primarily influenced by the distance between the metallic modes. It has been widely accepted that the coupling between the metallic channels near a Chern insulator junction interface can be reduced as distance increases\cite{LaGasse2016}.

Now, we manipulate the beam splitting by controlling the slope of the potential step. By changing the parameter $\xi$ in Eq. (\ref{eq:bipolarcontinuum}) from 0 to finite, the potential profile becomes smoothly varying, where its slope is tunable via gate control. Figure \ref{fg:dephasing}(a) shows that a suppression of AB conductance oscillation occurs when the gradualness of the potential slope increases. Compared to AB oscillation for an abrupt potential step, for the $\xi=10~a_{0}$ case the splitting imbalance increases due to weaker coupling between the metallic channels, resulting in the reduced visibility. With further $\xi$ increases, the AB conductance oscillation almost disappears due to the very weak coupling between the metallic channels. By comparing Figs. \ref{fg:dephasing}(b) and (c), for $\xi=20~a_{0}$, the incoming mode totally propagates through one arm of the AB interferometry, so that no enclosed area is created within the single path. Therefore, the AB conductance oscillation and related beating characteristics can be expected to be measurable only when the potential step is sharply fabricated.

\section{Conclusion} \label{sec:conclusion}

In conclusion, our results provide possible guides to experimental confirmation of the mesoscpic transport phenomena in quantum Hall graphene with a p-n junction. Especially, the valley-isospin dependence of the conductance through the p-n junction in quantum Hall graphene can be observed if the edges of graphene sample are perfectly clean. On the other hand, the conductance oscillation due to intrinsic AB interferometry can be measured when the p-n junction is sufficiently sharp. By examining the suppression of the AB oscillations, required sharpness of the junction is about $\leq$4.2 nm for a given magnetic field strength of 30 T, which seems to be feasible at present or in near future. Moreover, lower junction sharpness may be required for lower magnetic fields, since the enclosed area by the interface channels effectively increases as magnetic field decreases.

\section*{acknowledgement}
This work was supported by project code (IBS-R024-D1) and research fund from Chosun University, 2017. 

\appendix

\section{Decoupling sublattices of graphene wavefuctions} \label{sec:decouple}

Dirac Hamiltonian of the system reads
\begin{widetext}
\begin{align}
v_{F}\left(\begin{array}{cc}0&\frac{\hbar}{\mathrm{i}}\frac{d}{dx}-\mathrm{i}\left(\hbar k_{y}-eBx\right)\\\frac{\hbar}{\mathrm{i}}\frac{d}{dx}+\mathrm{i}\left(\hbar k_{y}-eBx\right)&0\end{array}\right)\left(\begin{array}{c}\psi_{A}\\ \psi_{B}\end{array}\right)=\left(E-V\right)\left(\begin{array}{c}\psi_{A}\\ \psi_{B}\end{array}\right),
\end{align}
\end{widetext}
where the Landau gauge $\vec{A}=\left(0,-Bx,0\right)$ is chosen, $V = V \left(x\right)$ as given by Eqs. (\ref{eq:bipolarcontinuum_a}) and (\ref{eq:bipolarcontinuum}) . For simplicity, let us treat $V$ as a constant $V_{0}$ and $-V_{0}$ in n- and p-doped regions, respectively. The above Dirac equation actually indicates two first-order differential equations as below:
\begin{align}
\hbar v_{F}\left[-\mathrm{i}\frac{d}{dx}-\mathrm{i}\left(k_{y}-\frac{eBx}{\hbar}\right)\right]\psi_{B}&=\left(E-V\right)\psi_{A},\label{eq:DiracA}\\
\hbar v_{F}\left[-\mathrm{i}\frac{d}{dx}+\mathrm{i}\left(k_{y}-\frac{eBx}{\hbar}\right)\right]\psi_{A}&=\left(E-V\right)\psi_{B},\label{eq:DiracB}
\end{align}
where $\psi_{A}$ and $\psi_{B}$ are coupled to each other via the Dirac equation. The wavefunction can be analytically solved by decoupling them. Here, a convenient way of decoupling is follows.

Since the Dirac equation leads to the relationship between sublattices, we have
\begin{align}
\psi_{A}&=\frac{\hbar v_{F}}{E-V}\left[-\mathrm{i}\frac{d}{dx}-\mathrm{i}\left(k_{y}-\frac{eBx}{\hbar}\right)\right]\psi_{B},\label{eq:psiA}\\
\psi_{B}&=\frac{\hbar v_{F}}{E-V}\left[-\mathrm{i}\frac{d}{dx}+\mathrm{i}\left(k_{y}-\frac{eBx}{\hbar}\right)\right]\psi_{A}.\label{eq:psiB}
\end{align}
Substituting Eq. (\ref{eq:psiB}) to Eq. (\ref{eq:DiracA}), we obtain
\begin{widetext}
\begin{align}
\left(\frac{E-V}{\hbar v_{F}}\right)^{2}\psi_{A}&=\left[-\mathrm{i}\frac{d}{dx}-\mathrm{i}\left(k_{y}-\frac{eBx}{\hbar}\right)\right]\left[-\mathrm{i}\frac{d}{dx}+\mathrm{i}\left(k_{y}-\frac{eBx}{\hbar}\right)\right]\psi_{A}\nonumber\\
&=\left[-\frac{d^{2}}{dx^{2}}+\left(k_{y}-\frac{eBx}{\hbar}\right)^{2}-\frac{eB}{\hbar}\right]\psi_{A}.
\end{align}
\end{widetext}
Therefore, the resulting equation is now a second-order differential equation for $\psi_{A}$:
\begin{align}
\left[\frac{d^{2}}{dx^{2}}-\left(k_{y}-\frac{eBx}{\hbar}\right)^{2}+\frac{eB}{\hbar}+\left(\frac{E-V}{\hbar v_{F}}\right)^{2}\right]\psi_{A}=0. \label{eq:2ndA}
\end{align}
Note that the third term $eB/\hbar$ occurs as a result of $\left[p_{x}, eA_{y}\right] = −i\hbar eB$. Similarly, by substituting Eq. (\ref{eq:psiA}) to Eq. (\ref{eq:DiracB}), we get
\begin{align}
\left[\frac{d^{2}}{dx^{2}}-\left(k_{y}-\frac{eBx}{\hbar}\right)^{2}-\frac{eB}{\hbar}+\left(\frac{E-V}{\hbar v_{F}}\right)^{2}\right]\psi_{B}=0. \label{eq:2ndB}
\end{align}
It is noteworthy that there is a sign change of the third term depending on sublattices. In results, two sublattice-coupled first-order differential equations from the Dirac equation have been expressed as one second-order differential equation via the decoupling process, which are equivalent to each other. In fact, we have two individual second-order differential equations for each sublattice, so we firstly obtain $\psi_{A}$ from Eq. (\ref{eq:2ndA}), and then find $\psi_{B}$ from the relationship between them using Eq. (\ref{eq:DiracB}), or vice versa.

Equations (\ref{eq:2ndA}) and (\ref{eq:2ndB}) are reduced in the following equation:
\begin{align}
\left[\frac{d^{2}}{dx^{2}}-\left(k_{y}-\frac{eBx}{\hbar}\right)^{2}+\frac{\varsigma eB}{\hbar}+\left(\frac{E-V}{\hbar v_{F}}\right)^{2}\right]\psi_{A,B}=0.
\end{align}
By making it dimensionless, we finally have Eq. (\ref{eq:effHam}):
\begin{align}
\left[\frac{d^{2}}{dx^{2}}-\left(k_{y}-\frac{x}{2}\right)^{2}+\frac{\varsigma}{2}+\left(E-V\right)^{2}\right]\psi_{A,B}=0,
\end{align}
where $E/E_{0}\rightarrow E$, $V/E_{0}\rightarrow V$, $l_{B}^{2}\left(d^{2}/dx^{2}\right)\rightarrow d^{2}/dx^{2}$, $k_{y}l_{B}\rightarrow k_{y}$, and $x/l_{B}\rightarrow x$.

\section{Obtaining eigenvalues from analytic solutions} \label{sec:eveq}

Solving Dirac equations, we found analytic solutions of the system as presented in Eq. (\ref{eq:analsols}). Unlike conventional quantum mechanics governed by the Schr\"{o}dinger equation, in relativistic-like quantum mechanics for graphene, the Dirac equation gives rise to the distinct boundary of condition for wavefunction continuity since it is basically a first-order differential equation. Thus, it is satisfactory to only have wavefunctions that are continuous at the interface of the Chern insulator junction, although their first derivation is no longer needed to be continuous, i.e.:
\begin{align}
\lim_{\varepsilon\rightarrow 0}\left[\Psi\left(0+\varepsilon\right)-\Psi\left(0-\varepsilon\right)\right]=0.
\end{align}
Now, this boundary condition leads to the following equation:
\begin{align}
A\left(\begin{array}{cc}D_{\nu}\left(-2k_{y}\right)&-D_{\nu}\left(2k_{y}\right)\\\mathrm{i}\sqrt{\frac{\nu}{2}}D_{\nu-1}\left(-2k_{y}\right)&\mathrm{i}\sqrt{\frac{\nu}{2}}D_{\nu-1}\left(2k_{y}\right)\end{array}\right)=0.
\end{align}
In order to have nonzero $A$, the determinant of the given $2\times 2$ matrix must vanish, so we can numerically find eigenenergies to satisfy the following:
\begin{align}
\det{\left(\begin{array}{cc}D_{\nu}\left(-2k_{y}\right)&-D_{\nu}\left(2k_{y}\right)\\\mathrm{i}\sqrt{\frac{\nu}{2}}D_{\nu-1}\left(-2k_{y}\right)&\mathrm{i}\sqrt{\frac{\nu}{2}}D_{\nu-1}\left(2k_{y}\right)\end{array}\right)}=0.
\end{align}
Finally, we reach the transcendental equation consisting of parabolic cylinder functions:
\begin{align}
D_{\nu}\left(-2k_{y}\right)D_{\nu-1}\left(2k_{y}\right)+D_{\nu}\left(2k_{y}\right)D_{\nu-1}\left(-2k_{y}\right)=0.
\end{align}
The resulting equation gives rise to eigenvalues as a function of $k_{y}$ as presented in Fig. \ref{fg:repfigs2}(a) and (c).

\section{Random edge roughness} \label{sec:random}

If our system is considered as rectangular, the system is characterized by its width W and length L, as depicted in Fig. 1(a). With straight-cut edges, we have constant W over the length L. On the other hand, if we are interested in putting irregular changes in the width, random roughness is necessary to be introduced. In other words, we want to make the edge fluctuate over the length L. The randomly fluctuating width can be created by setting W as a linear combination of sinusoidal functions, i.e.:
\begin{align}
W=W_{0}+\delta W\left[\sin{\left(\gamma_{1}x\right)}+\sin{\left(\gamma_{2}x\right)}\right],
\end{align}
where $\gamma_{1}$ and $\gamma_{2}$ are randomly given values in a range $\left[0.1L : 0.15L\right]$ by a random number generator (built-in function of Python 3.5), and $W_{0}$ and $\delta W$ are the constant values. The width now fluctuates between $W- \delta W$ and $W + \delta W$.

Since we put two individual random number generators for $\gamma_{1}$ and $\gamma_{2}$, the periods of the two sine functions are believed to be independent and different from each other. Thus, we expect that these functions are incommensurate resulting in an irregular shape of the edges in $−L/2 < x < L/2$. We accurately set the value of $\delta W$ to make the root-mean-square of the fluctuation about 0.53 $a_{0}$. Such a fluctuation amplitude value means that the fluctuation of the rough edge can cover one-atom-thick changes in the width $W$. Plus, every time we run a simulation, $\gamma_{1}$ and $\gamma_{2}$ are randomly distributed. Therefore, we run the simulation 100 times and take the ensemble average over the 100 simulation sets.

\bibliography{ABEGra}

\end{document}